\documentstyle[aps,twocolumn,psfig,floats]{revtex}

\begin{document}

\twocolumn[\hsize\textwidth\columnwidth\hsize\csname
@twocolumnfalse\endcsname
\title{
Quantum Monte Carlo calculations of H$_2$ dissociation on Si(001)}
\author{Claudia Filippi$^1$, Sorcha B. Healy$^2$, P. Kratzer$^3$, E. Pehlke$^4$,
and M. Scheffler$^3$}
\address{
$^1$Universiteit Leiden, Instituut Lorentz,  Niels Bohrweg 2, Leiden, NL-2333
CA, The Netherlands\\
$^2$Physics Department, National University of Ireland, Cork, Ireland\\
$^3$Fritz-Haber-Institut der Max-Planck-Gesellschaft, Faradayweg 4-6, D-14195
Berlin-Dahlem, Germany\\
$^4$Institut f\"ur Laser und Plasmaphysik, Universit\"at Essen, 45117
Essen, Germany}
\date{\today}
\maketitle

\begin{abstract}
We present quantum Monte Carlo calculations for various reaction pathways
of H$_2$ with Si(001), using large model clusters of the surface.
We obtain reaction energies and energy barriers noticeably higher than those
from approximate exchange-correlation functionals.  In improvement over
previous studies, our adsorption barriers closely agree with experimental data.
For desorption, the calculations give barriers for conventional pathways in
excess of the presently accepted experimental value, and pinpoint the role
of coverage effects and  desorption from steps.
\end{abstract}
\vskip2pc]

The dissociative adsorption of molecular hydrogen on the Si(001) surface
has become a paradigm in the study of adsorption systems. Despite its apparent
simplicity, more than a decade of extensive experimental and theoretical
investigations have not clarified fundamental aspects of the
chemical reaction of H$_2$ with this surface.

\begin{figure}[bht]
\noindent
\hspace*{0.3cm}\psfig{width=7.5cm,angle=0,figure=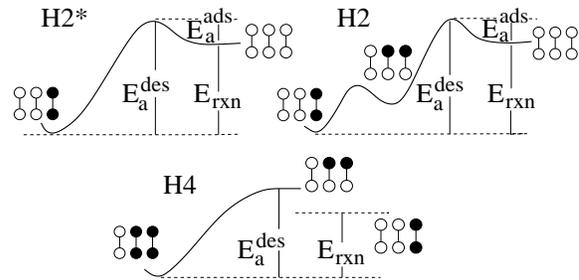}
\vspace{1.5ex}
\caption[]{Intra-dimer (H2*) and inter-dimer mechanisms at low (H2) and
high (H4) coverages. The surface configuration along the pathway is schematically
shown: A circle represents a Si atom and a filled circle Si--H.
E$^{\rm ads}_{\rm a}$, E$^{\rm des}_{\rm a}$ and E$_{\rm rxn}$ are the adsorption,
desorption and reaction energies, respectively.}
\label{fig1}
\end{figure}

Many of the experimental observations are hard to reconcile in a unified picture:
the sticking probability for dissociative adsorption of H$_2$ on the clean
surface is very small at room temperature suggesting a high adsorption
barrier; sticking increases dramatically with higher
surface temperatures~\cite{Arrhenius_stick}.
On the other hand, the nearly thermally distributed kinetic energy
of desorbing molecules has lead researchers to the conclusion that
the molecules have transversed almost no adsorption
barrier~\cite{noEkin}.
Microscopically, these observations were originally interpreted in terms of
an intra-dimer mechanism, where the hydrogen molecule interacts with one
single dimer of the Si(001) surface~\cite{pre-pairing}.
However, very recent experiments have pointed to additional mechanisms
involving not just a single dimer but nearby dimers~\cite{Zim,Heinz}.
The existence of highly reactive pathways was first demonstrated on
steps~\cite{Kratzer} or H-precovered surfaces~\cite{Hofer1}, and evidence
that H$_2$ reacts with two adjacent dimers has also now been given for the
clean surface~\cite{Hofer2}.
In Fig.~\ref{fig1}, the intra-dimer (H2*) and two inter-dimer pathways at
different coverages (H2 and H4) are schematically shown.

Theoretically, density-functional theory (DFT) calculations performed
on intra-dimer and inter-dimer mechanisms have lead to limited agreement with
experiments.
While correctly predicting the existence of a barrier-less H4
inter-dimer reaction path at high coverages~\cite{Pehlke},
previous DFT slab calculations yielded an adsorption barrier
for the low-coverage H2* and H2 pathways too low to explain the small
sticking coefficient observed at low temperatures~\cite{Hofer1}.
Desorption barriers from DFT obtained within the generalized gradient
approximation (GGA) were also generally lower than the experimental value
(2.5 eV, Ref.~\cite{Edes}).
Additional evidence for a possible inadequacy of DFT-GGA to describe
this reaction comes from comparison with highly correlated quantum
chemistry calculations for small cluster models of the
surface: For the intra-dimer pathway, these methods obtain
values for the desorption barrier that are at large variance with
the DFT-GGA value~\cite{Nachtigall,Carter,Jing}.

In this Letter, we use quantum Monte Carlo (QMC) techniques on large cluster
models of the surface to accurately compute the reaction energetics of H$_2$
on Si(001) for the intra-dimer and inter-dimer mechanisms, as well as
adsorption at steps.
The reliability of a calculation depends both on the level of theory at which
electronic correlations are treated, as well as on the geometrical model used to
describe the system. Compared to other theoretical approaches, QMC offers the
advantage that accurate reaction energetics can be computed for relatively large
systems.
Our calculations predict reaction energies and barriers noticeably higher than
those obtained in DFT with the local density approximation or commonly used
GGAs.  Even though distinctly lower, the energetics determined with the
B3LYP~\cite{B3LYP} hybrid functional are the only ones
in fair agreement with the QMC results.
When compared to experiments, the QMC adsorption barriers for the intra-dimer
(H2*) and the low-coverage inter-dimer (H2) pathways can explain why the sticking
coefficient on the clean Si(001) surface is so small at low temperatures.
Moreover, the QMC desorption barriers represent an important input for the
interpretation of experimental results.
\begin{table*}[tb]
\caption[]{
Adsorption (E$^{\rm ads}_{\rm a}$), desorption (E$^{\rm des}_{\rm a}$)
and reaction (E$_{\rm rxn}$) energies in eV for H$_2$/Si(001) via the
intra-dimer H2* mechanism (see Fig.~\ref{fig1}), calculated within PW91, B3LYP and QMC.
Zero-point energies (ZPE) are not included. }
\label{table1}
\begin{tabular}{lccc|ccc|ccc}
& \multicolumn{3}{c|}{PW91} & \multicolumn{3}{c|}{B3LYP} & \multicolumn{3}{c}{QMC} \\
&  E$^{\rm ads}_{\rm a}$ & E$^{\rm des}_{\rm a}$ & E$_{\rm rxn}$ &
   E$^{\rm ads}_{\rm a}$ & E$^{\rm des}_{\rm a}$ & E$_{\rm rxn}$ &
   E$^{\rm ads}_{\rm a}$ & E$^{\rm des}_{\rm a}$ & E$_{\rm rxn}$
\\
\hline
Si$_{9}$H$_{12}$  & 0.69    & 2.86    & 2.17
                  & 0.90    & 3.40    & 2.50
                  & 1.01$\pm$0.06 & 3.65$\pm$0.06 & 2.64$\pm$0.06 \\
Si$_{15}$H$_{16}$ & 0.56    & 2.65    & 2.09
                  & 0.71    & 3.20    & 2.49
                  & 0.98$\pm$0.05 & 3.52$\pm$0.07 & 2.54$\pm$0.06 \\
Si$_{21}$H$_{20}$ & 0.32    & 2.31    & 1.99
                  & 0.56    & 2.90    & 2.35
                  & 0.61$\pm$0.05 & 3.11$\pm$0.05 & 2.49$\pm$0.05 \\
Si$_{27}$H$_{24}$ & 0.37    & 2.35    & 1.98
                  & 0.57    & 2.91    & 2.33
                  & 0.49$\pm$0.13 & 3.03$\pm$0.13 & 2.54$\pm$0.13
\end{tabular}
\end{table*}

{\it Computational methods.}
We employ slabs as well as clusters to mimic the Si(001) surface.
The surface can be appropriately modeled with clusters containing only a single
row of dimers~\cite{Healy}: interactions are negligible between neighboring
dimer rows, while substantial between dimers in the same row.
Such clusters with one, two, three, and four dimers are Si$_9$H$_{12}$,
Si$_{15}$H$_{16}$, Si$_{21}$H$_{20}$, and Si$_{27}$H$_{24}$. They represent a
four layer cut of the Si(001) surface with all but the surface atoms terminated
with hydrogens to passivate dangling bonds.  Two separate clusters are constructed
to model the clean surface and the surface after adsorption of a hydrogen molecule;
the transition state (TS) connecting the the two configurations is then determined.

As a starting point, plane-wave pseudopotential calculations within DFT were
performed to obtain geometries and total energies of both the cluster and slab models.
For details on the construction and the geometry of the clusters, see
Ref.~\onlinecite{Penev}.
We optimized all geometries using the PW91 functional~\cite{PW91},
which gives a good description of the structural properties of Si
(lattice constant error $<1$\%).
Moreover, for cluster models of H$_2$/Si(001), the geometries optimized using
PW91 and the hybrid functional B3LYP were found to be very similar~\cite{Steckel},
and the energetics of the reaction on both sets of geometries essentially the same
within B3LYP (as shown below, B3LYP gives energies very close to our accurate
QMC results). It is therefore a sound procedure to use QMC on geometries
obtained from PW91 calculations to assess whether a more accurate treatment
of electronic correlation can change the physical picture.

We calculate the energetics for various reaction pathways of H$_2$ on the
Si(001) using B3LYP and QMC for the cluster geometries, and PW91 for both
clusters and slab.
The many-body wave function used in QMC is of the form given in
Ref.~\cite{Filippi} (modified to deal with pseudo-atoms):
\begin{eqnarray*}
\Psi=\sum_{n} d_n D_n^{\uparrow} D_n^{\downarrow}
\prod_{\alpha i j}J\left(r_{ij},r_{i\alpha},r_{j\alpha}\right)\,.
\end{eqnarray*}
${\rm D}^\uparrow_n$ and ${\rm D}^\downarrow_n$ are Slater determinants of
single particle orbitals for the up and down electrons, respectively, and the
orbitals are represented using atomic Gaussian basis~\cite{basis}.  The Jastrow
factor correlates pairs of electrons {\it i} and {\it j} with each other, and
with every nucleus $\alpha$, and different Jastrow factors are used to describe
the correlation with a hydrogen and a silicon atom.
The determinantal part of the wave function is generated within Hartree-Fock or MCSCF,
using the quantum chemistry package GAMESS~\cite{GAMESS}.  The parameters in the
Jastrow factor are optimized within QMC using the variance minimization
method~\cite{Umrigar_vmin} and the accuracy of the wave function tested at
the variational level.
The wave function is then used in diffusion Monte Carlo (DMC), which produces
the best energy within the fixed-node approximation (i.e. the lowest-energy state
with the same nodes as the trial wave function)~\cite{DMC}.
All QMC results presented are from DMC calculations.

{\it Intra-dimer mechanism.} In the H2* pathway, the hydrogen
molecule dissociatively adsorbs on the same Si dimer through an asymmetric TS.
Multi-reference configuration interaction (MRCI)
calculations~\cite{Jing} on Si$_9$H$_{12}$ yield adsorption and desorption
barriers which are 0.3 and 0.8~eV higher than the corresponding PW91 values
listed in Table~\ref{table1}.
Given the magnitude of the discrepancy, it is important to ascertain whether
this difference will persist in going from the one-dimer model to a more
realistic representation of the surface.
QMC gives energies in very good agreement with the MRCI results for
Si$_9$H$_{12}$ and, unlike traditional quantum chemistry methods, can be
applied to larger models of the surface to study the convergence of the reaction
energies with cluster size and access accurate estimates for the real surface.

In Table~\ref{table1}, we list the PW91, B3LYP and QMC adsorption, desorption
and reaction energies via the intra-dimer mechanism for the clusters with one,
two, three, and four surface dimers.
The QMC energies are consistently higher than the PW91 values for all
clusters: The QMC adsorption, desorption and reaction energies are above the
PW91 values by about 0.3, 0.8 and 0.5 eV, respectively.
Therefore, the corrections to the PW91 values due to the incorrect treatment of
electronic correlation are indeed significant, and, interestingly, they do not
show a noticeable dependence on the size of the cluster.
The QMC results for Si$_{27}$H$_{24}$ have a large errorbar but also
demonstrate the smooth QMC convergence with system size.
B3LYP represents a significant improvement upon PW91, giving
energies which are much closer to our QMC results, and only lower by
about 0.05, 0.2, and 0.15 eV for Si$_{21}$H$_{20}$.
This is in accordance with earlier studies using the B3LYP functional
for the Si--H system~\cite{Nachtigall,Steckel,Okamoto}.

\begin{table}[htb]
\caption[]{Adsorption, desorption and reaction energies in eV per molecule for
H$_2$/Si(001) via the H2 and H4 mechanisms (see Fig.~\ref{fig1}).
The Si$_{27}$H$_{24}$ model cluster of the surface is used.}
\label{table2}
\begin{tabular}{llll}
&  E$^{\rm ads}_{\rm a}$ & E$^{\rm des}_{\rm a}$ & E$_{\rm rxn}$\\[.1ex]
\hline\\[-2ex]
H2 mechanism & & & \\
PW91       & 0.26     & 2.24     & 1.99     \\
B3LYP      & 0.54     & 2.87     & 2.33     \\
QMC        & 0.59$\pm 0.09$  & 3.11$\pm 0.09$  & 2.52$\pm 0.09$  \\[1ex]
H4 mechanism & & & \\
PW91       & 0.00     & 2.46     & 2.13     \\
B3LYP      & 0.00     & 2.91     & 2.50     \\
QMC        & 0.19$\pm 0.14$ & 3.18$\pm 0.12$ & 2.61$\pm 0.11$ \\
\end{tabular}
\end{table}

{\it Inter-dimer H2 and H4 mechanisms.} In the H2 pathway, the hydrogen
molecule dissociates over two clean neighboring Si dimers, yielding a cis
configuration with two hydrogens bound to two Si atoms at the same side of
the dimers.
For the H4 mechanism, the adsorption occurs on two neighboring Si dimers which
are both already covered on the same side with hydrogens (see Fig.~\ref{fig1}).
The configuration after adsorption consists of two fully covered Si dimers.
The PW91, B3LYP and QMC adsorption, desorption and reaction energies for the H2
and H4 pathways are presented in Table~\ref{table2}. All calculations were
performed on the four-dimer cluster Si$_{27}$H$_{24}$ and the reaction occurs
on the two central dimers.
For both mechanisms, the results show the same trends as for the intra-dimer
pathway: PW91 significantly underestimates reaction energies and barriers, and
B3LYP is much closer to the QMC results than PW91.
For the H4 mechanism, PW91 and B3LYP predict no adsorption barrier.
Within the statistical error, we find this to remain true also in the QMC
calculation.
To explore a possible QMC adsorption barrier, we computed the QMC energies
for nine geometries along the path connecting the configuration of four
hydrogens on two neighboring dimers with the configuration of two hydrogens
in the cis configuration plus a desorbed hydrogen molecule.
However, the cis configuration
does {\em not} constitute the energetically lowest adsorption geometry
of two hydrogens atoms, and thus occurs only scarcely.
The pairing energy, that is the energy difference between the
cis configuration and two hydrogens at the same Si dimer,
is found to be 0.34, 0.42 and 0.54$\pm 0.07$ eV  in PW91, B3LYP and QMC,
respectively~\cite{cis_extend}.

{\it Discussion and conclusions.} In addition to accessing the
reliability of approximate DFT in describing H-Si bonded systems, we can use
our accurate QMC energetics to understand the physics of the
interaction of hydrogen with the Si(001) surface.
In order to compare with the available experimental results for the
H$_2$/Si(001) system, we need to extrapolate the QMC energies to the
infinite-system limit and include the zero-point energy corrections
($\Delta$ZPE).
To compute the extrapolated QMC results, we start from the PW91 infinite-system
limit, that is the PW91 slab energy, and add a correction for the inaccurate
treatment of electronic exchange and correlation estimated from the cluster calculation as
$\Delta$E$_{\rm corr} = {\rm E}^{\rm QMC}_{\rm cluster}-{\rm E}^{\rm PW91}_{\rm cluster}$.
After including the ZPE contribution, the final QMC energy is
${\rm E}^{\rm QMC}={\rm E}^{\rm PW91}_{\rm slab}+({\rm E}^{\rm QMC}_{\rm
cluster}- {\rm E}^{\rm PW91}_{\rm cluster})+\Delta{\rm ZPE}$.
In Table~\ref{table3}, the extrapolated QMC energies are compared with the
experimental results.

\noindent
\begin{table}[tb]
\caption[]{Extrapolated QMC adsorption, desorption and reaction energies
in eV for the H2*, H2 and H4 mechanisms (see text).
$^a$Ref.\onlinecite{Steckel} (identical $\Delta$ZPE assumed for all mechanisms);
$^b$Ref.~\onlinecite{Hofer1}; $^c$Ref.~\onlinecite{Edes}; $^d$Ref.~\onlinecite{Erxn}.}
\label{table3}

\begin{tabular}{lr@{.}lr@{.}lr@{.}l}
& \multicolumn{2}{c}{E$^{\rm ads}_{\rm a}$}
& \multicolumn{2}{c}{E$^{\rm des}_{\rm a}$}
& \multicolumn{2}{c}{E$_{\rm rxn}$} \\[.1ex]
\hline\\[-2ex]
\multicolumn{7}{l}{H2* mechanism}\\[.5ex]
E$^{\rm PW91}_{\rm slab}$   &    0&37     &    2&27     &    1&90   \\
$\Delta$ZPE$^a$  & $+$0&09     & $-$0&11     & $-$0&20   \\
$\Delta$E$_{\rm corr}$      & $+$0&29$\pm 0.05$  & $+$0&80$\pm 0.05$  & $+$0&50$\pm 0.05$ \\[1ex]
E$^{\rm QMC}$               &    0&75$\pm 0.05$  &    2&96$\pm 0.05$  &    2&20$\pm 0.05$ \\
Expt.                       &  $>$0&6$^b$ &    2&5$\pm 0.1^c$
                            &    1&9$\pm 0.3^d$  \\[2ex]
\multicolumn{7}{l}{H2 mechanism}\\[.5ex]
E$^{\rm PW91}_{\rm slab}$   &    0&20     &    2&15     &    1&95   \\
$\Delta$ZPE$^a$             & $+$0&09     & $-$0&11     & $-$0&20   \\
$\Delta$E$_{\rm corr}$      & $+$0&34$\pm 0.09$  & $+$0&87$\pm 0.09$  & $+$0&53$\pm 0.09$ \\[1ex]
E$^{\rm QMC}$               &    0&63$\pm 0.09$  &    2&91$\pm 0.09$  &    2&28$\pm 0.09$ \\
Expt.                       & $>$0&6$^b$  &    2&5$\pm 0.1^c$
                            &    1&9$\pm 0.3^d$  \\[2ex]
\multicolumn{7}{l}{H4 mechanism}\\[.5ex]
E$^{\rm PW91}_{\rm slab}$   &    0&00     &    2&32     &    2&01   \\
$\Delta$ZPE$^a$             &    \multicolumn{2}{c}{N/A} & $-$0&20     & $-$0&20   \\
$\Delta$E$_{\rm corr}$      & $+$0&19$\pm 0.14$ & $+$0&72$\pm 0.12$ & $+$0&48$\pm 0.11$\\[1ex]
E$^{\rm QMC}$               &    0&19$\pm 0.14$ &    2&84$\pm 0.12$ &    2&29$\pm 0.11$\\
Expt.                       &    0&00$^b$ &    2&5$\pm 0.1^c$
                            &    \multicolumn{2}{c}{N/A}
\end{tabular}
\end{table}

The PW91 slab reaction energies and barriers are significantly lower than the
experimental results and, for E$^{\rm des}_{\rm a}$ and E$_{\rm rxn}$, the discrepancy
is even amplified when the negative ZPE's are added.
As discussed above, the QMC energetics for the clusters are higher than the
corresponding PW91 values for all the mechanisms, so the correlation
corrections are positive and the extrapolated QMC results significantly higher
than the original PW91 slab energies. The final reaction energies are compatible
with experiments and the adsorption barriers are in much better
agreement with the experimental results than the PW91 barriers.
The larger adsorption barriers for the intra-dimer and the low-coverage H2
inter-dimer pathways can finally explain why the sticking coefficient on the
clean Si(001) surface is so small at low temperatures. Moreover, the high
reactivity
for adsorption via the H4 mechanism is corroborated by the QMC results
which are consistent with a barrier-less pathway.
Possibly, the dramatic increase in sticking probability with surface temperature is
partly due to this pathway: if hydrogen is already present on the surface,
an increase in surface temperature leads to an increased number of cis configurations,
thus creating barrier-less adsorption sites for sticking.

Concerning desorption, the picture is more complicated.
Judging from the extrapolated QMC barriers, none of the studied mechanisms
appears to be compatible with the desorption energy of 2.5$\pm 0.1$~eV observed
in temperature-programmed desorption experiments~\cite{Edes}.
However, such experiments could have been affected by a small concentration of
surface imperfections. To exemplify this possibility, we looked at the
adsorption/desorption of H$_2$ at the D$_{\rm B}$ step edge which is modeled as
a Si$_{28}$H$_{28}$ cluster constructed from previous slab calculations~\cite{Kratzer}.
Experimentally, a small adsorption barrier of 0.09$\pm 0.01$ eV is observed while PW91
predicts no adsorption barrier, so PW91 desorption and reaction energies are the
same.  Here, we compute only the B3LYP and QMC reaction energies for
Si$_{28}$H$_{28}$,
which are equal to 2.54 and 2.81$\pm 0.07$ eV, respectively. The extrapolated QMC
value is 2.61$\pm 0.07$ eV, significantly lower than the other desorption barriers,
and hence the presence of a small number of steps could dominate the desorption
yield.  Our calculations suggest that the hitherto accepted experimental value
of the desorption barrier should be referred to desorption from steps or defects,
rather than to one of the mechanisms discussed for ideal Si(001) surfaces.
For experiments where contributions from surface imperfections are
carefully avoided (e.g.\ by heating the surface only very locally),
the QMC calculations predict a slight preference for the H4 mechanism.
While the lack of an adsorption barrier along the H4 pathway can explain the
low kinetic energy of the desorbing hydrogen molecules, one should also keep
in mind the observed vibrational excitation in desorption~\cite{vibrations},
indicating that some molecules desorb on a pathway with an adsorption
barrier.  We conclude that several mechanisms contribute to desorption,
whose relative importance depends sensitively on temperature, coverage, and
surface perfection.

In this Letter, we presented accurate QMC calculations for various pathways
of adsorption/desorption of H$_2$ from Si(001), using large cluster models
of the surface. For intra-dimer and inter-dimer pathways, we find that PW91
significantly underestimates reaction energies and barriers, while B3LYP
yields energetics in much better agreement with the QMC values.
Caution should therefore be used when employing PW91 or other GGAs in the
study of Si--H systems. Finally, the QMC adsorption barriers are in close
agreement with experimental values while the results for desorption call
for further experimental studies of the activation energy and its
dependence on coverage and surface perfection.


\begin{references}
\bibitem{basis}
The Gaussian basis sets are (10$s$10$p$1$d$)/[3$s$3$p$1$d$] for silicon
and (10$s$1$p$)/[3$s$1$p$] for hydrogen, and are optimized
at the HF level for the Si$_9$H$_{12}$ cluster.
\bibitem{cis_extend}
The PW91 slab value for the pairing energy is 0.31 eV.
\end{references}
\end{document}